\documentclass{aastex631}

\usepackage{amsmath, amssymb}
\usepackage{amsmath, bm, mathtools, cancel, empheq, ulem, mathrsfs, natbib}
\usepackage{bm, graphicx}
\usepackage[counterclockwise]{rotating}
\usepackage{multirow, threeparttable}

\newcommand{\vecb}{\boldsymbol{B}}
\newcommand{\vecv}{\boldsymbol{v}}
\newcommand{\bpol}{\vecb_{\rm{pol}}}
\newcommand{\vpol}{\vecv_{\rm{pol}}}

\newcommand{\sn}[2]{#1\times10^{#2}}

\newcommand{\ri}{r_{\rm{min}}}
\newcommand{\ro}{r_{\rm{max}}}
\newcommand{\rbcz}{r_{\rm{bcz}}}
\newcommand{\rov}{r_{\rm{ov}}}
\newcommand{\rtach}{r_{\rm{tach}}}

\newcommand{\rhoref}{\overline{\rho}}

\newcommand{\sref}{\overline{S}}

\newcommand{\prm}{{\rm{Pr_m}}}
\newcommand{\prt}{{\rm{Pr}}}

\newcommand{\prot}{P_{\rm{rot}}}

\newcommand{\rsun}{R_\odot}
\newcommand{\lsun}{L_\odot}

\newcommand{\tauin}{\tau_{\rm{in}}}

\newcommand{\tauv}{\tau_{\rm{visc}}}
\newcommand{\taumag}{\tau_{\rm{mag}}}
\newcommand{\taumagm}{\tau_{{\rm{mag}},m}}
\newcommand{\taumagmw}{\tau_{{\rm{mag}},m\omega}}

\newcommand{\av}[1]{\left\langle#1\right\rangle}
\newcommand{\avt}[1]{\left\langle#1\right\rangle_{\rm{t}}}

 \newcommand{\newtext}{}
 \newcommand{\nnewtext}[1]{#1}

\shorttitle{Solar Tachocline Confinement by Interior Dynamo Action}
\shortauthors{Matilsky et al.}

\begin{document}
	
\title{Confinement of the Solar Tachocline by Dynamo Action in the Radiative Interior} 

	\correspondingauthor{Loren I. Matilsky}
	\email{loren.matilsky@gmail.com}	
	\author[0000-0001-9001-6118]{Loren I. Matilsky}
	\affiliation{JILA \& Department of Astrophysical and Planetary Sciences, 
	University of Colorado Boulder, 
	Boulder, CO 80309-0440, USA}	

	\author[0000-0001-7612-6628]{Bradley W. Hindman}
	\affiliation{Department of Applied Mathematics,
	University of Colorado Boulder,
	Boulder, CO 80309-0526, USA}
	\affiliation{JILA \& Department of Astrophysical and Planetary Sciences,
	University of Colorado Boulder,
	Boulder, CO 80309-0440, USA}
	
		\author[0000-0003-1077-3368]{Nicholas A. Featherstone}
	   \affiliation{Southwest Research Institute,
	   	1050 Walnut Street Suite 400,
	   	 Boulder, CO 80302, USA}
   	 
		\author[0000-0002-3125-4463]{Catherine C. Blume}
	   \affiliation{JILA \& Department of Astrophysical and Planetary Sciences,
		University of Colorado Boulder,
		Boulder, CO 80309-0440, USA}
	
	\author[0000-0002-3125-4463]{Juri Toomre}
	\affiliation{JILA \& Department of Astrophysical and Planetary Sciences,
		University of Colorado Boulder,
		Boulder, CO 80309-0440, USA}

	\begin{abstract}
A major outstanding problem in solar physics is the confinement of the solar tachocline, the thin shear layer that separates nearly solid-body rotation in the radiative interior from strong differential rotation in the convection zone. Here, we present the first 3-D, global solar simulation \newtext{that displays a magnetically confined tachocline.} The non-axisymmetric magnetism is initially built in the convection zone and then diffusively imprints downward, \newtext{similar to the proposed fast magnetic confinement scenario by the Sun's cyclic dynamo field.} Additionally, the field is locally amplified throughout the radiative interior by vigorous horizontal motions that \newtext{seem to arise from a combination of equatorial Rossby waves and shear, magnetic, and buoyancy instabilities. Our work thus supports prior studies proposing dynamo action in the radiative interior, and suggests that horizontal motions could play a key role in driving this deep dynamo.}
	\end{abstract}

	
	
	\section{Introduction} \label{sec:intro}
  In the solar tachocline at the base of the convection zone, strong differential rotation ($\sim$30\% faster at the equator than at the poles) transitions to nearly solid-body rotation in the radiative interior (e.g., \citealt{Brown1989,Howe2000}). Helioseismic estimates of the tachocline's width lie around $0.05\rsun$ \citep{Howe2009}, where $\rsun\equiv\sn{6.96}{8}$ m is the solar radius. Because of this strong shear, the tachocline likely plays a central role in the solar dynamo. The ``interface" dynamo paradigm, in particular, holds that toroidal magnetism is primarily generated by the tachocline's shear (e.g., \citealt{Parker1993,Charbonneau1997}), and is then stored for long time intervals in the quiescent radiative interior (e.g., \citealt{Spruit1982,Parker1993,FerrizMas1994}).
  
  Without an opposing mechanism, inward radiative diffusion of the latitudinal temperature gradient in the convection zone is expected to drive meridional circulations in the stable layer that would have imprinted differential rotation deep into the interior by the current age of the Sun \citep{Spiegel1992}. To achieve a thin tachocline, the Sun must have a torque that forces solid-body rotation in the radiative interior and thus counters radiative spread. Several prevalent tachocline confinement scenarios postulate the origin of this torque.  In the ``fast hydrodynamic (HD) scenario" \newtext{(e.g., \citealt{Spiegel1992,Brun2017b,Cope2020})}, the torque is caused by the Reynolds stresses associated with primarily HD instabilities of the horizontal flows and is generated on the timescale of months to years. In the \newtext{``slow magnetohydrodynamic (MHD) scenario", or ``magnetic scenario" (e.g., \citealt{Rudiger1997,Gough1998,MacGregor1999})}, the torque is due to a primordial magnetic field and is generated on the timescale of radiative spread, namely, $\sim$$10^{11}$ years. Finally, in the ``fast MHD scenario," the torque comes from the cyclic dynamo magnetic field in the convection zone (timescale of $\sim$22 years) imprinting diffusively downward to a skin depth, similar to the skin effect for AC currents in a conductor \newtext{(e.g., \citealt{ForgcsDajka2001,ForgcsDajka2002,ForgcsDajka2004,Barnabe2017}).}
  
  Many theoretical studies have characterized the instabilities believed to cause a fast HD scenario (e.g., \citealt{Charbonneau1999,Gilman2014,Garaud2020}) and have examined how an assumed primordial or cycling magnetic field might cause a slow or fast MHD scenario (e.g., \citealt{Garaud2002,AcevedoArreguin2013,Barnabe2017,Wood2018}). For global, 3-D simulations, computationally tractable values of the thermal Prandtl number and buoyancy frequency do not permit substantial radiative spread  (e.g., \citealt{AcevedoArreguin2013, Wood2012}); instead, the tachocline spreads viscously. Transient tachoclines have been included in prior global dynamo simulations by implementing a very small viscosity below the convection zone \newtext{(e.g., \citealt{Brun2011, Passos2014, Guerrero2016a})}. The tachocline still spreads inward slowly, but for the timescale on which the simulation is run, it is effectively stationary and its influence on the dynamo can be assessed. 
  
  Here, we present two 3-D, global, nonlinear simulations of a rotating solar-like star---an HD case and an MHD case---that include a radiative interior coupled to an outer convection zone. In the HD case, the differential rotation viscously imprints throughout the entire radiative interior. In the MHD case, by contrast, dynamo action creates a cycling, non-axisymmetric magnetic field whose torque enforces solid-body rotation in the radiative interior and maintains a statistically steady tachocline. The magnetism in the radiative interior arises both from diffusive imprinting of field from the overlying convection zone (similar to the fast MHD confinement scenario), and also from local inductive amplification by strong horizontal motions. This dynamo action occurs even below the convective overshoot layer, \newtext{a phenomenon also suggested by prior mean-field calculations (e.g., \citealt{Dikpati2001b, Spruit2002, Bonanno2013}) and explored in global dynamo simulations (e.g., \citealt{Racine2011, Lawson2015}). In our simulation, the horizontal motions in the radiative interior are due to equatorially confined (equatorial) Rossby waves \citep{Gizon2020a} and possibly shear, magnetic, and buoyancy instabilities as well.} 
  
\section{Numerical Experiment}
We use the Rayleigh code \citep{Featherstone2016a,Matsui2016,Featherstone2021} to evolve the anelastic fluid equations (e.g., \citealt{Gilman1981}) in a rotating spherical shell that spans $\ri=0.491\rsun$ to $\ro=0.947\rsun$. We use spherical coordinates: $r$ (radius), $\theta$ (colatitude), and $\phi$ (azimuth angle). \newtext{We denote the vector velocity and magnetic fields by $\vecv$ and $\vecb$, respectively.} The background stellar structure is hydrostatic, spherically symmetric, and time-independent. The background entropy gradient enforces strong convective stability in the radiative interior and weak convective instability in the convection zone. The transition between stability and instability nominally occurs at $r_0\equiv0.719\rsun$. More details on the thermodynamic state are given in Appendix \ref{ap:thermo}.

Our shell covers approximately equal thickness in both the convection zone and radiative interior, corresponding to the top $\sim$2.1 density scale heights of the Sun's radiative interior and the bottom 3 density scale heights of the convection zone. Our grid resolution is $N_\theta = 384$ and $N_\phi = 768$ in the horizontal directions (the maximum spherical-harmonic degree after dealiasing is $\ell_{\rm{max}}=255$). We use three stacked Chebyshev domains in the vertical direction (each with 64 grid points) with boundaries at $(0.491, 0.669, 0.719, 0.947) \rsun$. The two internal boundaries maximize grid resolution at the transition from stability to instability.


As in \citet{Matilsky2020a, Matilsky2021}, the simulations rotate at three times the solar Carrington rate ($\Omega_0=3\Omega_\odot$, where $\Omega_\odot=2.87\times10^{-6}\ \rm{rad\ s^{-1}}$). Rotating faster than the Sun is required to avoid the ``anti-solar" states associated with the simulations' overestimation of the fluctuating velocities at large scales (e.g., \citealt{OMara2016}). The frame rotation frequency is $\Omega_0/2\pi=1370$ nHz and the frame rotation period is $\prot\equiv2\pi/\Omega_0=8.45$ days. A solar luminosity $\lsun\equiv3.85\times10^{33}\ \rm{erg\ s^{-1}}$ is driven through the convection zone via a fixed internal-heating profile and is removed at the outer boundary via thermal conduction. 

At the top of the domain, $\nu(r)=\kappa(r)=5.00\times10^{12}\ \rm{cm^2\ s^{-1}}$ and $\eta(r)=1.25\times10^{12}\ \rm{cm^2\ s^{-1}}$, where $\nu(r)$, $\kappa(r)$, and $\eta(r)$ are the momentum, thermal, and magnetic diffusivities, respectively. All diffusivity profiles increase with radius like $\rhoref(r)^{-1/2}$, where $\rhoref(r)$ is the background density. At both boundaries, we use stress-free and impenetrability conditions on the velocity, potential-field-matching conditions on the magnetic fields, and fixed-conductive-flux conditions (e.g., \citealt{Matilsky2020b, Anders2020}) on the entropy.

The convection is initialized by weakly perturbing the thermal field randomly throughout the entire shell. The MHD case is additionally initialized by weakly perturbing the magnetic field randomly throughout the convection zone only. We define the HD case's ``equilibrated state'' as the time interval during which the kinetic energy in the radiative interior is statistically steady. For the MHD case's equilibrated state, we additionally require that the magnetic energy in the radiative interior is statistically steady. We define the viscous and magnetic diffusion times across the radiative interior to be $[r_0-\ri]^2/\nu_0 = 295\ \prot$ and $[r_0-\ri]^2/\eta_0 =1180\ \prot$, respectively, where the ``0" subscript indicates the value of the diffusivity at $r_0$. The HD case was run in its equilibrated state for 7810 $\prot$ (26.5 viscous diffusion times) and the MHD case for $12500\ \prot$ (10.6 magnetic diffusion times).   

The convection zone (defined to be the region where the convective heat transport is positive) has a base $\rbcz$ that is set by the balance of radial energy fluxes in equilibrium. Convective downflows overshoot into a thin layer within the stable region, the top of which is $\rbcz$ and the base of which is $\rov$ (defined to be the location below which there is negligible convective heat transport). For the MHD case, $\rbcz=0.729\rsun$ and $\rov=0.710\rsun$. We define the radiative interior as the layer spanning $\ri$ to $\rov$. Calculation of $\rov$ and $\rbcz$ for each case is done explicitly in Appendix \ref{ap:thermo}. The non-dimensional parameters characterizing each case are given in Appendix \ref{ap:nond}. 
  
 \section{Steady-State Solid-Body Rotation in the Radiative Interior}
\begin{figure}
 	\centering
 	\includegraphics{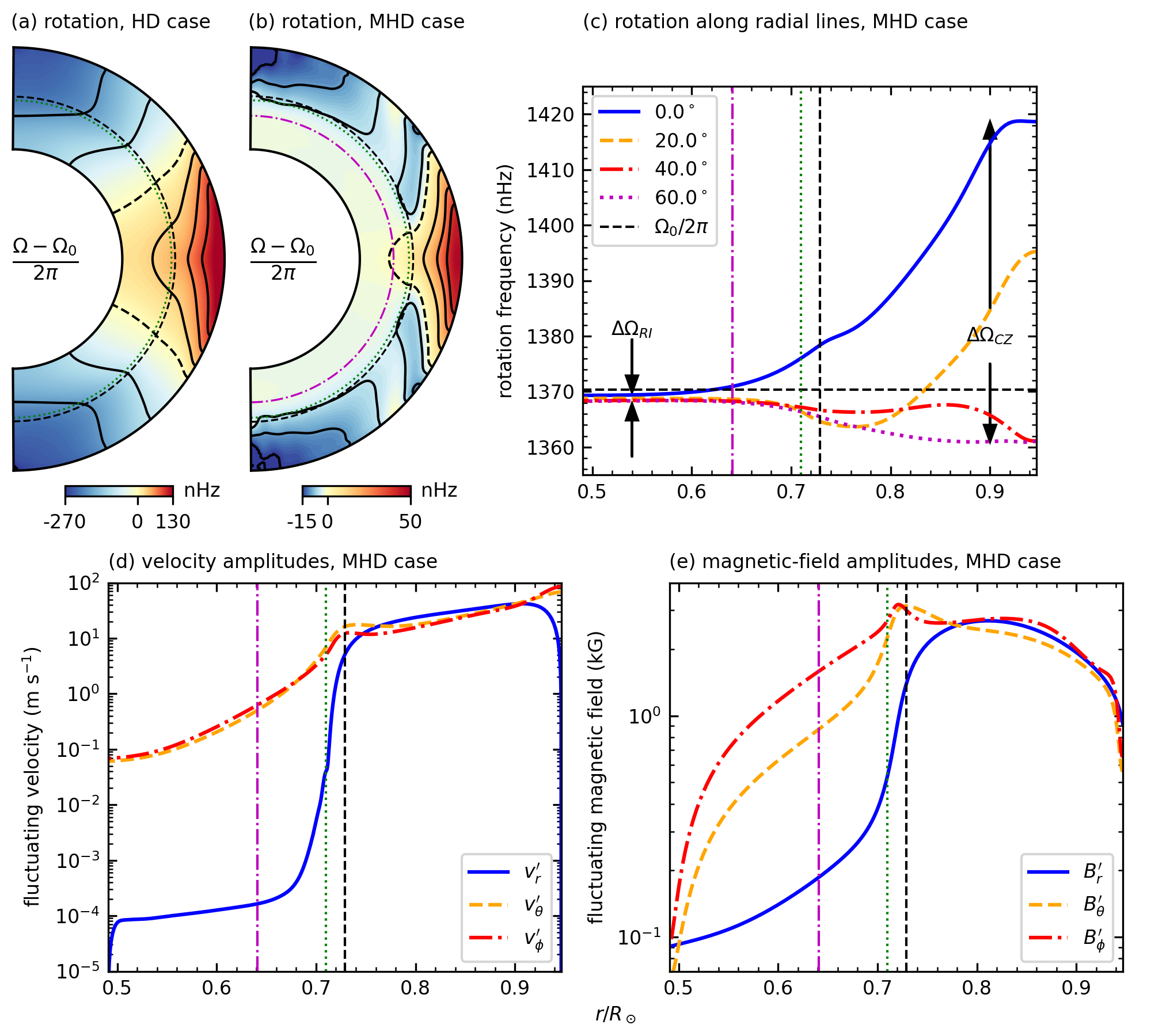}
 	\caption{Radiative interior forced into solid-body rotation. (a) Contours of isorotation, $(\Omega-\Omega_0)/2\pi=\text{constant}$, in the HD case, plotted in the meridional plane. Negative values are normalized separately from positive values. There are three equally-spaced positive and negative contours each (solid contours). The zero contour, $\Omega=\Omega_0$, is dashed. (b) Like (a), but for the MHD case. (c) Rotation frequency $\Omega/2\pi$ at various latitudes along radial lines in the MHD case. The rotation contrasts in the convection zone and radiative interior are marked by vertical arrows; the frame rotation frequency is marked by the horizontal dashed line. (d) \newtext{Fluctuating velocity amplitudes $v_\alpha^\prime\equiv \av{(v_\alpha - \av{v_\alpha})^2}_{\rm{sph}}^{1/2}$, where the index $\alpha$ denotes $r$, $\theta$, or $\phi$ and the temporal average in $\av{\cdots}_{\rm{sph}}$ is taken over the full equilibrated state. Each component $v_r^\prime$, $v_\theta^\prime$, and $v_\phi^\prime$ is plotted separately. (e) Same as (d), but for the fluctuating magnetic-field amplitudes $B_\alpha^\prime$ (defined similarly to $v_\alpha^\prime$)}. In this figure and those that follow, the dashed black, dotted green, and dash-dotted magenta curves refer to $\rbcz$, $\rov$, and $\rtach$, respectively.}
 	\label{fig:flows}
 \end{figure}

 We define the zonally and temporally averaged rotation rate as $\Omega(r,\theta)\equiv \Omega_0 + \avt{v_\phi}/r\sin{\theta}$ and the rotation frequency as $\Omega/2\pi$. Throughout the text, angular brackets with no subscript denote an instantaneous zonal average, whereas a ``t" subscript denotes a combined temporal and zonal average, and a ``sph" subscript denotes a combined temporal and spherical-surface average. \newtext{The time interval over which temporal averages are taken will be explicitly stated in the text; for the rotation rate $\Omega$, we average over the full equilibrated state.} Figures \ref{fig:flows}(a, b) show the rotation-frequency profiles for both simulations in the meridional plane. In the HD case, the differential rotation has viscously imprinted throughout the entire radiative interior. In the MHD case, however, the radiative interior has nearly solid-body rotation. 
 
 Figure \ref{fig:flows}(c) shows the rotation profile in the MHD case as a function of radius for various latitudes. We define the radially varying latitudinal rotation contrast $\Delta\Omega(r)$ as the difference in rotation rate between the equator and $60^\circ$ latitude at a fixed radius. In the HD case, $\Delta\Omega(r)/\Omega_0\sim0.2$ throughout the whole shell. The MHD case has $\Delta\Omega(r)/\Omega_0=4.2\times10^{-2}\equiv\Delta\Omega_{\rm{CZ}}/\Omega_0$ at the top of the convection zone and $\Delta\Omega(r)/\Omega_0=7.6\times10^{-4}\equiv\Delta\Omega_{\rm{RI}}/\Omega_0$ at the bottom of the radiative interior. We define the base of the tachocline, $\rtach=0.641\rsun$, to be the radial location where $\Delta\Omega(r)$ has dropped by a factor of 20 from its value at the top of the convection zone \newtext{(i.e., $\Delta\Omega_{\rm{CZ}}$)} and call the layer spanning $\rtach$ to $\rbcz$ the MHD case's tachocline.
 
 The velocity and magnetic-field amplitudes for the MHD case are shown in Figures \ref{fig:flows}(d, e). Below the overshoot layer, the vertical components of the fluctuating velocity, $\vecv^\prime\equiv\vecv-\av{\vecv}$, and fluctuating magnetic field, $\vecb^\prime\equiv\vecb-\av{\vecb}$, are small compared to the horizontal components. Due to the stable stratification, $v_r^\prime$ falls by $\sim$2 orders of magnitude over the overshoot layer and by $\sim$5 orders of magnitude over the whole radiative interior.
 
\begin{figure}
 	\centering
 	\includegraphics{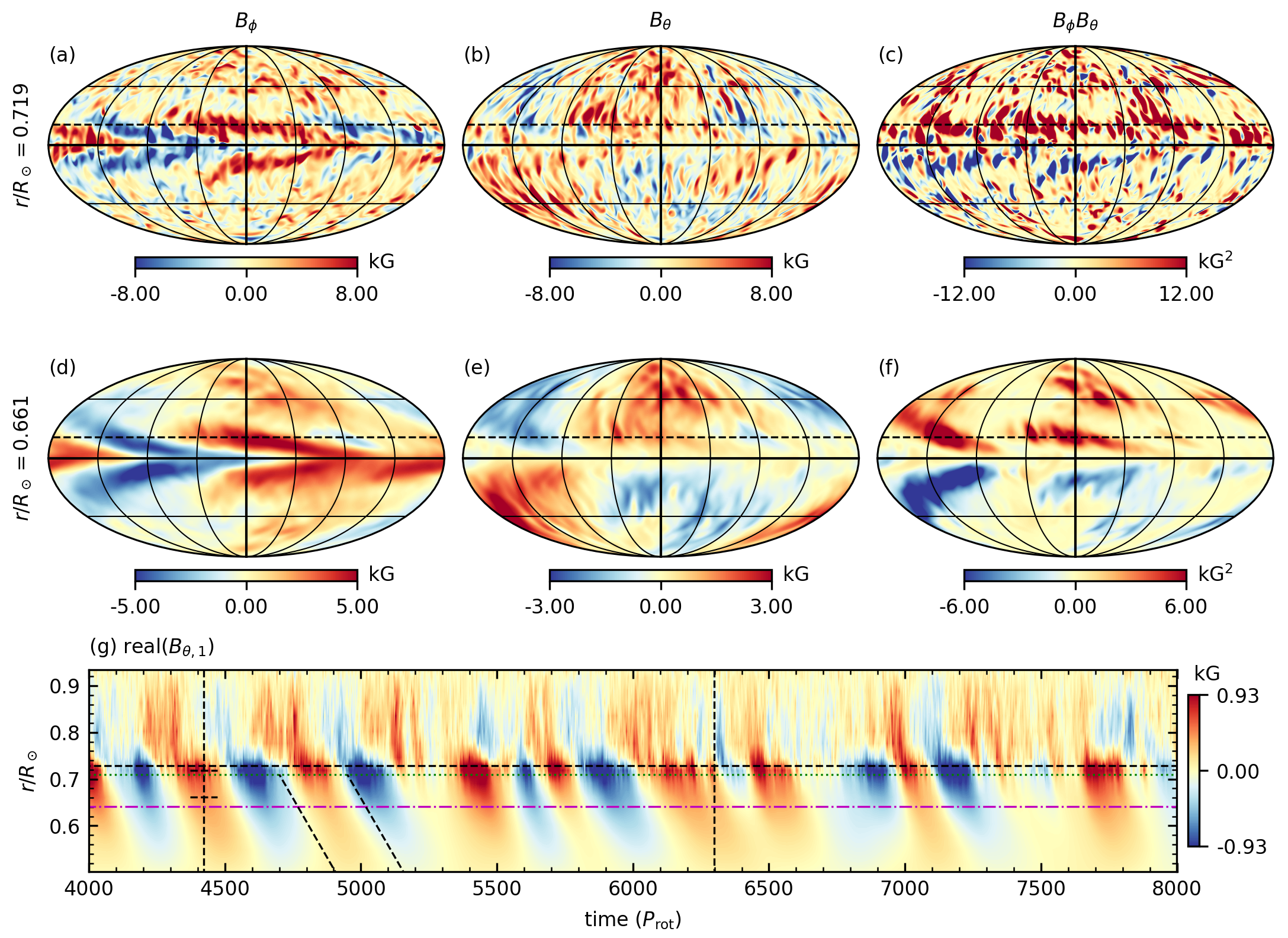}
 	\caption{Non-axisymmetric, cycling dynamo. Snapshots of the horizontal magnetic fields \newtext{($B_\phi$, $B_\theta$, and their product, $B_\phi B_\theta$) are shown at $t=4422\prot$ for (a)--(c) the overshoot layer and (d)--(f) the tachocline.} Each field is plotted in full Mollweide view, latitudes and longitudes are marked every $45^\circ$ by solid curves, and $15^\circ$ latitude is marked by a dashed line. (g) The real part of the $m=1$ component of $B_\theta$ \newtext{(i.e., $\rm{real}(B_{\theta,1})$, where $B_{\theta,1}\equiv\av{B_\theta\exp{(-i\phi)}}$)} plotted as a function of time and radius at $15^\circ$ latitude. The vertical dashed lines denote the  time interval considered in \newtext{Section \ref{sec:torque} and} Figure \ref{fig:torque}. The two ticks on the leftmost vertical line show the depths sampled by the Mollweides. The diagonal dashed lines show the speed at which diffusion would imprint the oscillating field downward from the base of the overshoot layer according to the skin effect. \newtext{The skin depth is $\sim$$0.08\rsun$, on par with the distance strong $B_{\theta,1}$ extends below the overshoot layer.} }
 	\label{fig:moll}
 \end{figure}

 The non-axisymmetric magnetic field in the MHD case is composed mainly of azimuthal orders $m=1$ and $m=2$. Figures \ref{fig:moll}(a--f) show snapshots of the horizontal field components ($B_\phi$, $B_\theta$, \newtext{and their product}) at two different depths. At both depths, the same large-scale structure is apparent in each field component, though it is significantly smoother and more coherent in the deeper layer. \newtext{Notably, the product $B_\phi B_\theta$ (which is important in computing the magnetic torque discussed at length in Section \ref{sec:torque}) is positive in the North and negative in the South at both depths, giving poleward angular-momentum flux that tends to enforce solid-body rotation.} The topology of the fields is similar to the ``partial wreaths" we identified in convection-zone only dynamo simulations \citep{Matilsky2020a,Matilsky2020c}. In that work, a regularly cycling dynamo composed of full magnetic wreaths (strong $m=0$ signature) destabilized into two partial wreaths (strong $m=1$ signature). The partial wreaths were essentially stationary in a properly chosen rotating frame and reversed their polarity (sign of the $m=0$ component) through in-place amplitude modulation. The polarity-reversal time varied, forming a quasi-cyclic dynamo with multiple frequency components. 
 
 Figure \ref{fig:moll}(g) shows the real part of the $m=1$ component of $B_\theta$ (two partial wreaths) as a function of time and radius. The partial wreaths appear first in the convection zone and then move downward into the overshoot layer, where they are significantly amplified, before finally penetrating deep into the radiative interior. The other field components ($B_r$ and $B_\phi$) also behave this way, as do the $m=2$ structures. From Figure \ref{fig:moll}(g), one quasi-regular dynamo cycle occurs every $P_{\rm{dyn}}\sim500\ \prot$. Furthermore, the partial wreaths migrate downward at approximately the same speed predicted by the skin effect, namely $v_{\rm{skin}}\equiv\sqrt{2\eta_{\rm{ov}}\omega_{\rm{dyn}}}$, where $\omega_{\rm{dyn}}/2\pi\equiv1/P_{\rm{dyn}}=2.74$ nHz and $\eta_{\rm{ov}}$ is the value of the magnetic diffusivity at the base of the overshoot layer $r=\rov$. The skin depth is $\delta_{\rm{skin}}\equiv\sqrt{2\eta_{\rm{ov}}/\omega_{\rm{dyn}}}= 0.08\rsun$, which allows the transport of field significantly below the overshoot layer.
 
\section{Dynamical Maintenance of Solid-Body Rotation}\label{sec:torque}
The MHD case's tachocline is statistically steady. \newtext{In particular, although the instantaneous rotation rate $\av{v_\phi}/r\sin\theta$ does vary slightly with time from the profile for $\Omega$ shown in Figures \ref{fig:flows}(b,c), these variations are no more than $\sim$3 nHz at any given point below the convection zone.} There is thus good temporally and zonally averaged torque balance. \newtext{For our anelastic approximation, the torque equation takes the form (see, e.g., \citealt{Miesch2011}; their Equation (A4)):}
\begin{align}
	&\underbrace{\nabla\cdot\left[\rhoref\nu r^2\sin^2\theta\nabla(\avt{v_\phi}/r\sin\theta)\right]}_{\text{viscous torque}\ \equiv\ \tauv}+  \underbrace{(1/4\pi)\nabla\cdot \left[r\sin\theta \avt{ B_\phi\bpol}\right]}_{{\text{magnetic torque}\ \equiv\ \taumag}} \underbrace{-\ \nabla\cdot\left[\rhoref r\sin\theta\avt{v_\phi\vpol}\right]}_{\text{inertial torque}\ \equiv\ \tauin} = 0, 
	\label{eq:torque}
\end{align}
where $\bpol\equiv B_r\hat{\boldsymbol{e}}_r + B_\theta\hat{\boldsymbol{e}}_\theta$  \newtext{is the poloidal magnetic field and $\vpol\equiv v_r\hat{\boldsymbol{e}}_r + v_\theta\hat{\boldsymbol{e}}_\theta$ is the poloidal velocity field. Equation \eqref{eq:torque} expresses torque balance under steady-state conditions, wherein the local angular momentum is statistically steady.  In this Section, temporal averages are taken over the sub-interval of our MHD case, $4422\prot$ to $6299\prot$ (vertical dashed lines in Figure \ref{fig:moll}(g)), encompassing about four dynamo cycles.} 

 \newtext{Because the dynamo magnetic fields in our simulation are non-axisymmetric and cycling, we decompose the vector magnetic field $\vecb$ into its constituent azimuthal order ($m$) and frequency ($\omega$) components ($\vecb_{m\omega}$), and consider the separate contributions to the magnetic torque from each $m$ and $\omega$. Each component $\vecb_{m\omega}$ is the coefficient of a Fourier mode $\exp{(im\phi-i\omega t)}$ that moves in azimuth angle with phase velocity $\omega/m$. We normalize these coefficients such that $\sum_m\sum_\omega|\vecb_{m\omega}|^2=\avt{ |\vecb|^2}$. We then define the magnetic torque $\taumagmw$ from a given  $\vecb_{m\omega}$:}
\begin{align}
	\taumagmw\equiv \frac{1}{4\pi}\nabla\cdot [r\sin\theta B_{\phi,m\omega}^*\vecb_{{\rm{pol}},m\omega}],
	\label{eq:magtorquemw}
\end{align}
where the asterisk denotes the complex conjugate. We further define 
\begin{align}\label{eq:magtorquem}
\taumagm\equiv\sum_\omega\taumagmw.
\end{align}
  \newtext{Clearly, the sum over all components $\taumagmw$ gives back the full magnetic torque: $\sum_m\taumagm=\sum_m\sum_\omega\taumagmw=\taumag$. }
 \begin{figure}
	\centering
	\includegraphics{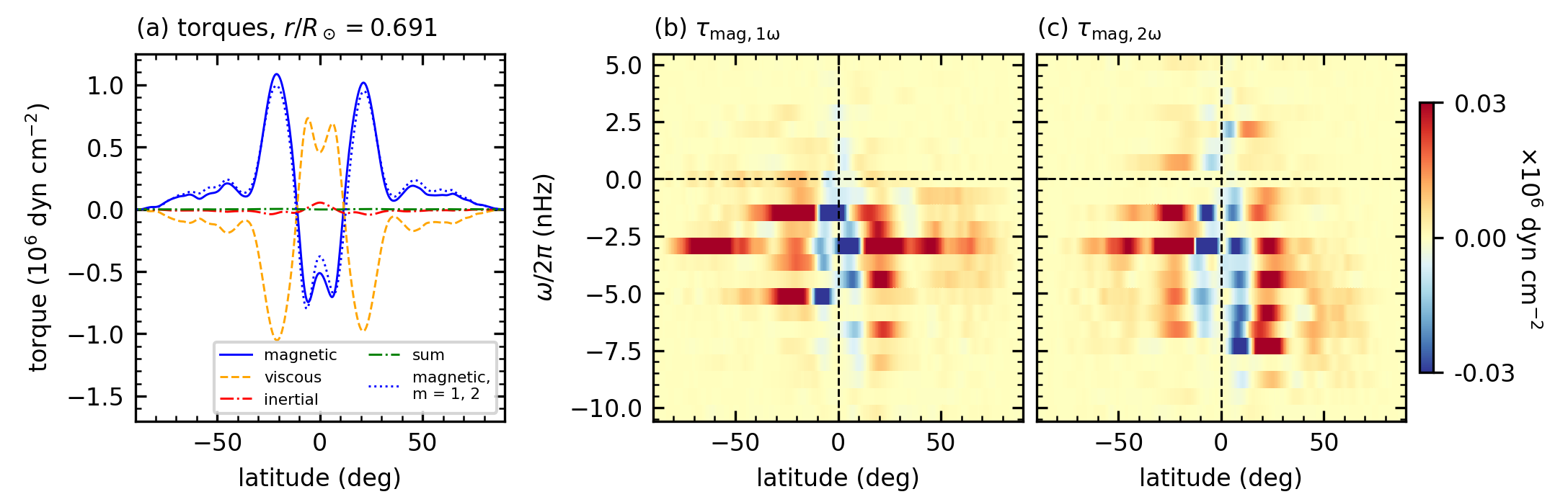}
	\caption{Magnetic torque maintaining the tachocline, as sampled at $r/\rsun=0.691$. (a) \newtext{The three torques defined in Equation \eqref{eq:torque}, and their sum, as functions of latitude in the tachocline. Also shown: the magnetic torque from just the $m=1,2$ fields ($\tau_{\rm{mag,1}} + \tau_{\rm{mag,2}}$).}  (b) and (c) The frequency-decomposed contributions to $\tau_{\rm{mag,1}}$ and $\tau_{\rm{mag,2}}$, respectively. For the time interval  $4422\prot$ to $6299\prot$, the frequency resolution is $0.73$ nHz and the Nyquist frequency is $5$,$000$ nHz.}
	\label{fig:torque}
\end{figure}

The torque balance in the tachocline is shown in Figure \ref{fig:torque}(a). Viscosity works to imprint the differential rotation from above by spinning up the equator and slowing down the mid-latitude regions, but is halted by the magnetic torque.  \newtext{The contribution from the inertial torque (i.e., Reynolds stresses and transport by the meridional circulation) is negligible. The sum of all torques (dashed black curve) is very close to zero, confirming the fact that the MHD case has reached a statistically steady state. We also plot the magnetic torque from just the $m=1$ and $m=2$ components of the magnetic field as the dotted blue curve (see Equation \eqref{eq:magtorquem}). We see that the magnetic torque is almost entirely due to just these two components. Physically, this happens because the fields $B_\phi$ and $\bpol$ (each of which are individually dominated by their $m=1,2$ components) \nnewtext{have a zonal phase difference that is small enough such} that their product, which is zonally averaged in computing $\taumag$, always has a strong $m=0$ component (recall Figures \ref{fig:moll}c,f).} 

\newtext{We next consider the latitudinal variation of the magnetic torque due to the different frequency components of the non-axisymmetric ($m=1,2$) magnetic fields. We first examine the $\tau_{\rm{mag,1\omega}}$ components in Figure \ref{fig:torque}(b) and the $\tau_{\rm{mag,2\omega}}$ components in Figure \ref{fig:torque}(c). Remarkably, at any given frequency, the latitudinal profile of torque (i.e., the values associated with a given horizontal strip in Figures \ref{fig:torque}b,c) is basically the same, and resembles the magnetic-torque curve in Figure \ref{fig:torque}(a). Physically, this means that \textit{each frequency component of the magnetic field separately opposes the viscous spread of the tachocline.} In other words, the magnetic field in the fast magnetic confinement scenario may not be restricted to its originally proposed form of an axisymmetric field having the single frequency component associated with the 22-year cycle (e.g., \citealt{ForgcsDajka2001, Barnabe2017}). Instead, Figure \ref{fig:torque} suggests that even a predominantly non-axisymmetric magnetic field with complicated, only quasi-periodic, cycling behavior can efficiently confine the solar tachocline.}

\section{Non-Axisymmetric Ferraro's Law}\label{sec:ferraro}
Ferraro's law of isorotation was originally stated for temporally steady and axisymmetric magnetic fields in stellar radiative interiors: ``Contours of isorotation tend to fall along poloidal magnetic field lines" \citep{Ferraro1937}. The argument behind this law (\newtext{which we here closely paraphrase from \citealt{Mestel1987}})  is perhaps even more relevant in the non-axisymmetric, cycling context. Differential rotation bends poloidal magnetic field lines to produce a toroidal field through mean shear. \newtext{The toroidal component of the MHD induction equation is}
\begin{subequations}\label{eq:indp}
	\begin{align}
	\frac{\partial B_\phi} {\partial t}&=\underbrace{[\nabla\times(\vecv\times\vecb)]_\phi}_{\text{induction}}\  \underbrace{-\ [\nabla\times(\eta(r)\nabla\times\vecb)]_\phi}_{\text{diffusion}}\label{eq:indp1}\\
	\text{with}\ \ \ \ \ [\nabla\times(\vecv\times\vecb)]_\phi &=
	\underbrace{r\sin\theta\vecb_{\rm{pol}}\cdot\nabla \left(\frac{v_\phi}{r\sin\theta}\right)}_{\text{shear}}
	\ \underbrace{-\ B_\phi\left[\frac{\partial v_r}{\partial r} +\frac{1}{r}\frac{\partial v_\theta}{\partial\theta} + \frac{v_r}{r}\right]}_{\text{compression}} 
	 \ \underbrace{-\ \vecv\cdot \nabla B_\phi}_{\text{advection}}. \label{eq:indp2}
	\end{align}
\end{subequations}
 When the mean shear \newtext{($r\sin\theta\bpol\cdot\nabla\Omega$)} dominates \newtext{the other terms}, Equation \eqref{eq:indp} yields $B_\phi\approx(r\sin\theta\bpol\cdot\nabla\Omega)t$, \newtext{where $t$ measures the time from which the shear was imposed, and is assumed to be short enough that $\bpol$ remains effectively constant in time (i.e., for our MHD case, we consider $t\ll P_{\rm{dyn}}\approx500\prot$).}
 
  \newtext{From Equation \eqref{eq:torque}, the mean shear produces an instantaneous magnetic torque, \nnewtext{$\tau_{\rm{Ferraro}}\equiv\nabla\cdot\left[r^2\sin^2\theta \av{\bpol(\bpol\cdot\nabla\Omega)}\right](t/4\pi)$}, which \newtext{locally} tends to eliminate gradients in $\Omega$ parallel to $\bpol$. In the absence of other torques on the system, equilibrium then requires zero (temporally and zonally averaged) magnetic torque, or, excluding very special configurations for the rotation rate and poloidal field,} $\avt{(\bpol\cdot\nabla\Omega)\bpol} = 0$. For an axisymmetric, temporally steady poloidal magnetic field, this reduces to the original Ferraro's law, $\avt{\bpol}\cdot\nabla\Omega = 0$. However, as noted by \citet{Mestel1987}, a non-axisymmetric field \newtext{is more restrictive; since a zonally varying $\bpol$ efficiently eliminates shear along different directions at different longitudes, the magnetic torque from such a field tends to induce solid-body rotation. \nnewtext{We also note that the tendency toward solid-body rotation should be even stronger for a cycling dynamo, since $\bpol$ efficiently eliminates shear along different directions during different \textit{cycles} as well as at different longitudes}.}
  	
  \newtext{\citet{Mestel1987} showed that for both axisymmetric and non-axisymmetric fields, locally imposed shear of length-scale $D$ would be eliminated on the timescale $D/v_A$, where $v_A^2=\bpol^2/4\pi\rhoref$ is the squared Alfv\'en velocity associated with the \textit{poloidal} magnetic field. For our MHD case's tachocline, where $|\bpol|\approx1000\ \rm{G}$, $\rhoref\approx0.2\ \rm{g\ cm^{-3}}$, and $D\leq\ro-\rbcz\approx\sn{1.6}{10}$ cm (see Figure \ref{fig:flows}), we have $v_A\approx 6.3\ \rm{m\ s^{-1}}$ and $t=D/v_A\approx 35\prot$, \nnewtext{a timescale much smaller than our dynamo cycle period. This suggests that a process like non-axisymmetric Ferraro's law enforces solid-body rotation in our MHD case by eliminating any viscously imposed shear effectively instantaneously on the timescale of the cycling dynamo. We caution that the estimate $t\approx35\prot$ is not a universal constant for our MHD case. Rather, for any particular viscously imposed shear pertrubation, the exact $t$ will be locally determined by the amplitude of $\bpol$ and the length-scale of the shear perturbation. The actual magnetic torque of Figure \ref{fig:torque} thus cannot be computed directly from the above expression for $\tau_{\rm{Ferraro}}$. Nevertheless, a process like non-axisymmetric Ferraro's law should operate in our MHD case as long as (1) $B_\phi$ is produced predominantly via mean shear and (2) a significant $\bpol$ is sustained at all times.}}

\section{Magnetization of the Radiative Interior}\label{sec:dynamo}
\begin{figure}
	\centering
	\includegraphics{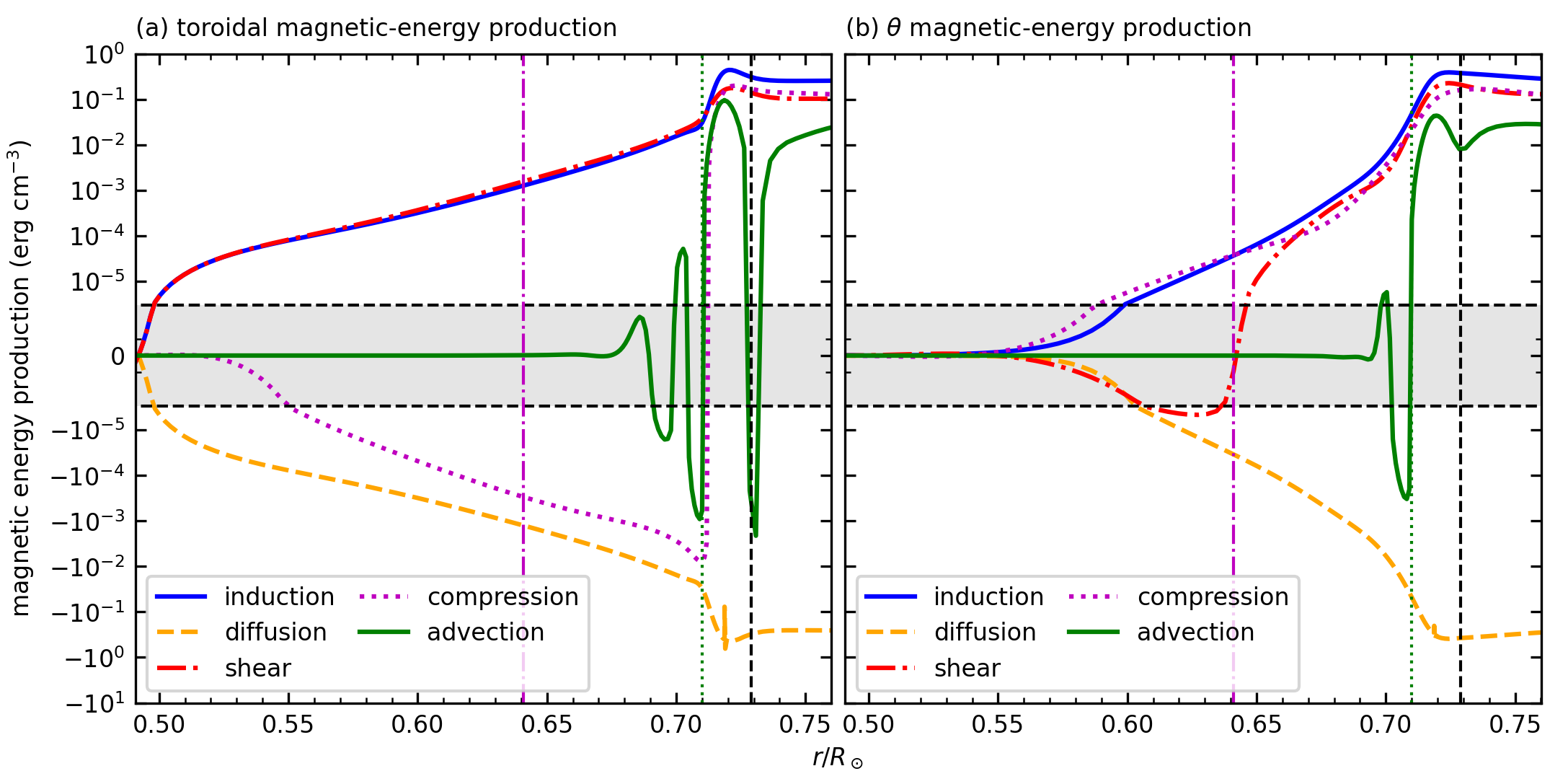}
	\caption{Dynamo action in the MHD case's radiative interior. Production rate of (a) toroidal magnetic energy $B_\phi^2/8\pi$ and (b) $\theta$ magnetic energy $B_\theta^2/8\pi$, plotted with respect to radius below $r/\rsun=0.760$. The terms labeled \newtext{on the right-hand sides of Equations \ref{eq:indp} and \ref{eq:indt} are multiplied by $B_\phi/4\pi$ and $B_\theta/4\pi$, respectively, and then averaged over spherical surfaces and in time over the equilibrated state to yield radial profiles}.  Each panel shares the same $x$- and $y$-axes. The $y$-axis is scaled logarithmically (unshaded region) for absolute values $>3\times10^{-6}\ \rm{erg\ cm^{-3}}$ and linearly (shaded region) otherwise.  }
	\label{fig:meprod}
\end{figure}

\newtext{For consistency with non-axisymmetric Ferraro's law, the toroidal magnetic field in our MHD case must be produced predominantly via mean shear. To verify this, we multiply Equation \eqref{eq:indp} by $B_\phi/4\pi$ and then average over spherical surfaces and time over the equilibrated state to form a balance between toroidal magnetic-energy production terms. These terms are plotted as the curves in Figure \ref{fig:meprod}(a). In the radiative interior, toroidal field is sustained mostly by the shear (dashed-dotted red curve) and dissipated diffusively. We have verified that at all points in the radiative interior, the energy production by \textit{mean} shear makes up at least 80\% of the total shear production. Note that \nnewtext{turbulent} magnetic \nnewtext{pumping} (the magnetic-energy production from radial advection \nnewtext{by the turbulent flow field}, which dominates the total advection shown as the solid green curves in Figure \ref{fig:meprod}(a)) has a role in transporting toroidal-field energy from the convection zone to the base of the overshoot layer. Below the overshoot layer, the advective transport is negligible, as it must be for incompressible horizontal flows.}

Figure \ref{fig:meprod}(a) shows that the MHD case's tachocline stems from strong $\bpol$ (which, considering Figure \ref{fig:flows}(e), is dominated by $B_\theta$), also consistent with Ferraro's law. Broadly speaking, $B_\theta$ in the radiative interior can arise either through inward diffusion of the convection zone's $B_\theta$ or through local inductive amplification. We have already seen from Figure \ref{fig:moll}(e) that the first of these mechanisms is significant. To isolate the second mechanism, we consider the $\theta$-component of the induction equation:
\begin{subequations}\label{eq:indt}
\begin{align}
		\frac{\partial B_\theta} {\partial t}&=\underbrace{[\nabla\times(\vecv\times\vecb)]_\theta}_{\text{induction}}\  \underbrace{-\ [\nabla\times(\eta(r)\nabla\times\vecb)]_\theta}_{\text{diffusion}}\label{eq:indt1}\\
	\text{with}\ \ \ \ \ [\nabla\times(\vecv\times\vecb)]_\theta &=  
	 \underbrace{r(B_r\hat{\boldsymbol{e}}_r+ B_\phi\hat{\boldsymbol{e}}_\phi)\cdot\nabla \left(\frac{v_\theta}{r}\right)}_{\text{shear}}\ 
	\underbrace{-\ B_\theta\left(\frac{1}{r\sin\theta}\frac{\partial v_\phi}{\partial\phi} + \frac{\cot\theta}{r} v_\theta+ \frac{\partial v_r}{\partial r} +\frac{v_r}{r}\right)}_{\text{compression}}
	 \ \underbrace{-\ \vecv\cdot \nabla B_\theta}_{\text{advection}}. \label{eq:indt2}	
\end{align}
\end{subequations}

The production of $\theta$ magnetic energy $B_\theta^2/4\pi$---$\av{\cdots}_{\rm{sph}}$ applied to the product of Equation \eqref{eq:indt} and $B_\theta/4\pi$---is shown in Figure \ref{fig:meprod}(b). On long timescales, $B_\theta$ is destroyed by diffusion and amplified inductively at all radii. The induction comes primarily from compression (the zonal squeezing of $B_\theta$; contributions from the $v_r$ terms are negligible) and shear (the tilting of radial and toroidal field into $B_\theta$). Like for $B_\phi$, magnetic pumping deposits $B_\theta$ from the convection zone to the base of the overshoot layer. \newtext{Note that the dynamo action in our radiative interior does not form a closed loop. $B_\phi$ is produced from $B_\theta$ via mean shear, but $B_\theta$ is produced mainly by compression, especially in the deeper layers. Compression simply amplifies the existing ``seed" $B_\theta$ that is originally transported diffusively from the partial wreaths in the convection zone. The partial wreaths themselves seem due to a non-axisymmetric $\alpha$$\Omega$-type dynamo, as we briefly explored in \citet{Matilsky2020a}.}

\section{Discussion}
Our MHD case represents \newtext{a 3-D, non-axisymmetric, quasi-cyclic version of the fast magnetic confinement scenario. In this modified scenario,} the poloidal field  penetrates downward diffusively and is locally amplified in the deep layers by induction. The original fast MHD scenario relies on a turbulently enhanced magnetic diffusivity to make the nominal solar-cycle skin depth (only $\sim$10 km) on par with the tachocline thickness \citep{ForgcsDajka2001}. However, the Sun's full dynamo field, which has modulations on many different timescales, like grand minima and the biennial oscillation (e.g., \citealt{Hathaway2015}), may extend deeper than this nominal skin depth. \newtext{If different frequency components of the solar dynamo can produce similar torques (as suggested by our Figure \ref{fig:torque}), then temporal variations other than the primary 22-year dynamo cycle could work to confine the solar tachocline.}



\begin{figure}
	\centering
	\includegraphics[width=4.75in]{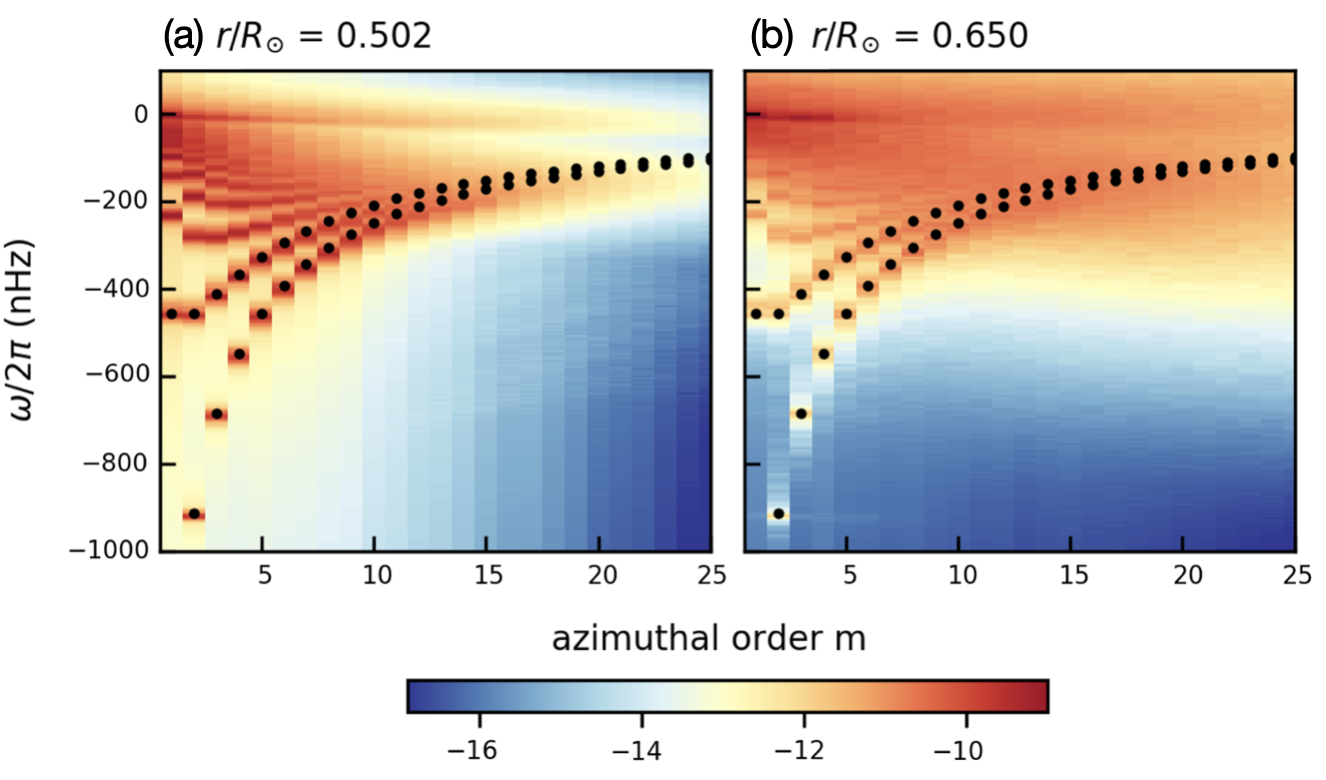}
	\caption{Rossby waves in the radiative interior. Power in the radial vorticity (integrated over latitude) as a function of $m$ and $\omega$ for (a) the deep interior and (b) the tachocline. For $\ell-m=0$ and $\ell-m=1$ (0 latitudinal nodes in the spherical harmonic and 1 node, respectively), the dispersion relation given by Equation \eqref{eq:disp} is plotted versus $m$ using black dots. The power has been averaged over eight realizations of the wavefield, each of length $206\prot$. For each realization, the frequency resolution is 5 nHz and the Nyquist frequency is 5,000 nHz.}
	\label{fig:rossby}
\end{figure}

\newtext{Furthermore, if the poloidal field can be inductively amplified locally, then an extremely enhanced value of the turbulent magnetic diffusivity may not be necessary. Even a small seed poloidal field could grow to amplitudes capable of tachocline confinement. The source of strong horizontal motions has not been identified definitively, though they frequently appear in the radiative interiors of global simulations (e.g., \citealt{Brun2011,Alvan2014,Lawson2015,Guerrero2016a,Bice2020,Bice2022}). The motions have been attributed to a combination of shear, magnetic, and buoyancy instabilities (e.g., \citealt{Gilman1997, Dikpati1999, Lawson2015, Gilman2018a}). In our case, the motions are primarily a superposition of many resonant equatorial Rossby waves \citep{Gizon2020a}.} Figure \ref{fig:rossby} shows the power in the radial vorticity with respect to $m$ and $\omega$ (summed over all latitudes). We overlay the theoretical dispersion relation for equatorial Rossby waves,
\begin{equation}
	\omega_{\ell m}=-\frac{2\Omega_0m}{\ell(\ell+1)},
	\label{eq:disp}
\end{equation}
where $\ell$ is the spherical-harmonic degree (e.g., \citealt{Zaqarashvili2021}). In the deep layers, the power in radial vorticity is closely aligned with Equation \eqref{eq:disp}, with small frequency shifts due to the effects of the differential rotation and cycling magnetic field. \newtext{At high $m$ and at low frequency, the ridges corresponding to different values of $\ell-m$ blend together, since the line widths of the power profiles become larger than the frequency spacing between the ridges.} Furthermore, Rossby waves in this frequency range would be close to the background rotation rate and significantly affected by the magnetic field, yielding critical-latitude, high-latitude, and MHD Rossby waves \citep{Gizon2020a,Zaqarashvili2021}. In the tachocline, our Rossby waves lie amidst a largely featureless background, and their low-frequency signature is stronger. This background may be due to overshooting plumes, which should impart significant vorticity to the deeper layers in a stochastic fashion \citep{Tobias1998}, \newtext{as well as the previously mentioned instabilities.}

In \citet{Gilman1969a}, it was shown that Rossby waves in the convection zone could achieve a complete dynamo loop. Studies have also shown that unstable Rossby waves likely exist in the solar tachocline and could produce poloidal magnetic field (e.g., \citealt{Charbonneau1999b,Gilman2018a,Zaqarashvili2021}). Many of these instabilities set in at $m=1$ and $m=2$ (e.g., \citealt{Garaud2001,Lawson2015,Gilman2018a}). In \citet{Charbonneau1999b}, the shear instability arises from critical-latitude Rossby waves, which appear when the frequency from Equation \eqref{eq:disp} is comparable to the latitudinal differential rotation contrast. These results are strongly suggestive of critical-latitude (potentially unstable) Rossby waves forming the low-frequency signature of power in Figure \ref{fig:rossby}. 

Equatorial and critical-latitude Rossby waves have been identified in the upper convection zone by helioseismic means \citep{Loptien2018,Gizon2020a}. The critical-latitude modes, in particular, may have a significant concentration of energy near the base of the convection zone \citep{Gizon2021}. They could thus produce strong horizontal motions at least down into the tachocline and our work suggests that such motions may extend even deeper. In summary, we offer two new perspectives on the solar interior. First, tachocline confinement is possible by a self-excited, 3-D dynamo magnetic field that is non-axisymmetric and has multiple cycling frequency components. Second, strong horizontal motions may amplify magnetic field locally through induction, even below the overshoot layer. Both perspectives suggest a dynamically active radiative interior and challenge its perceived role as a quiescent storage reservoir.

\begin{acknowledgments}
We thank M. Miesch, N. Brummell, P. Garaud, C. Bice, A. Brun, and K. Augustson for helpful discussions. L. Matilsky was primarily supported during this work by the Future Investigators in NASA Earth and Space Sciences Technology (FINESST) award 80NSSC19K1428. C. Blume was supported by a University of Colorado George Ellery Hale Graduate Fellowship. This research was primarily supported by NASA Heliophysics through grant 80NSSC18K1127, with additional support by NASA through grants 80NSSC18K1125, 80NSSC19K0267, 80NSSC17K0008, and 80NSSC20K0193. Computational resources were provided by the NASA High-End Computing (HEC) Program through the NASA Advanced Supercomputing (NAS) Division at Ames Research Center. Rayleigh is supported by the Computational Infrastructure for Geodynamics (CIG) through NSF awards NSF-0949446 and NSF-1550901. Input files and checkpoint snapshots for the HD and MHD cases are publicly accessible via Zenodo \citep{Matilsky2022}, with more extensive datasets available from the authors upon request.
\end{acknowledgments}

\appendix
\restartappendixnumbering
\section{Background thermodynamic state}\label{ap:thermo}
In our models, we employ a spherically symmetric, time-independent background state that represents the stable-to-unstable transition in the Sun that occurs at the base of the convection zone. We choose a simplified entropy-gradient profile $d\sref/dr$, which is zero in the convection zone, has a constant positive value in the stable layer, and has smooth matching in between:
\begin{equation}
	\frac{d\overline{S}}{dr} = \begin{cases}
		\sigma & r\leq r_0 - \delta\\
		\sigma\bigg{\{}1 - \Big{[}1 - \Big{(}\frac{r-r_0}{\delta}\Big{)}^2\Big{]}^2\bigg{\}} & r_0 - \delta < r < r_0\\
		0 & r\geq r_0,
	\end{cases}
\label{eq:dsdr}
\end{equation}
where $\sigma\equiv10^{-2}\ \rm{erg\ g^{-1}\ K^{-1}\ cm^{-1}}$ and $\delta\equiv0.05\rsun$. We choose a background gravitational-acceleration profile of 
\begin{equation}
	g(r)=\frac{GM_\odot}{r^2},
	\label{eq:grav}
\end{equation}
where $G=6.67\times10^{-8}\ \rm{dyn\ cm^2\ g^{-2}}$ is the gravitational constant and $M_\odot=1.99\times10^{33}\ \rm{g}$ is the mass of the Sun. We write the pressure, density, and temperature as $\overline{P}(r)$, $\overline{\rho}(r)$, and $\overline{T}(r)$, respectively. Hydrostatic balance and the ideal-gas condition yields
\begin{subequations}
	\begin{align}
		\overline{T}(r) =& -\exp{\left[\frac{\overline{S}(r)}{c_{\rm{p}}}\right]} \int_{r_0}^r \frac{g(x)}{c_{\rm{p}}} \exp{\left[-\frac{\overline{S}(x)}{c_{\rm{p}}}\right]}dx\nonumber\\
		&+ \overline{T}_0\exp{\left[\frac{\overline{S}(r)}{c_{\rm{p}}}\right]},\\
		\overline{P}(r)=\ &\overline{\rho}_0\mathcal{R}\overline{T}_0\exp\bigg{[}-\frac{\overline{S}(r)}{\mathcal{R}}\bigg{]}\bigg{[}\frac{\overline{T}(r)}{T_0}\bigg{]}^{\gamma/(\gamma - 1)},\\
		\text{and}\ \ \ \ \ \overline{\rho}(r)=\ &\overline{\rho}_0\exp\bigg{[}-\frac{\overline{S}(r)}{\mathcal{R}}\bigg{]}\bigg{[}\frac{\overline{T}(r)}{T_0}\bigg{]}^{1/(\gamma - 1)}.
	\end{align}
\label{eq:thermo}
\end{subequations}
Here, $c_{\rm{p}}={3.50}\times10^8\ \rm{erg\ g^{-1}\ K^{-1}}$ is the specific heat at constant pressure, $\gamma=1.67$ is the ratio of specific heats, and $\mathcal{R}=(\gamma -1)c_{\rm{p}}/\gamma$ is the gas constant. We choose $\overline{\rho}_0=0.181\ \rm{g\ cm^{-3}}$ and $\overline{T}_0={2.11}\times10^{6}\ \rm{K}$, consistent with solar models \citep{ChristensenDalsgaard1996}, and $\overline{S}(r_0)=0$. \newtext{We do not solve for the radiation field explicitly and instead include the heating by radiation via a fixed internal heating profile $Q(r)$, which is chosen} to occupy the convection zone only:
\begin{equation}
	Q(r)= \alpha\tanh{\left[\frac{r-r_0}{\delta_{\rm{heat}}}\right]}[\overline{P}(r)-\overline{P}(\ro)],
	\label{eq:heat}
\end{equation}
where $\delta_{\rm{heat}}=0.03\rsun$ and the constant $\alpha$ is chosen so the volume integral of $Q(r)$ over the whole shell is the solar luminosity. In the convection zone, the reference state is nearly identical to our prior work and closely resembles the standard solar ``model S" \citep{Featherstone2016a, Matilsky2020b, Hindman2020}.

In keeping with past work \citep{Brun2017}, we define the convection zone (and hence its base $\rbcz$) to be the region in which the convective heat flux (or enthalpy flux $F_e$) is positive. Similarly, we define the base of the overshoot layer $\rov$ as the location below which $F_e$ is negative but very close to zero (we choose a tolerance of 5\% the minimum value of $F_e$ in the overshoot layer). Though the nominal transition between stability and instability occurs at $r_0$, convective heat transport moves the base of the convection zone slightly upward to $\rbcz>r_0$. Convective downflows overshoot into a thin layer within the stable region. The base of this overshoot layer (defined to be the location below which there is negligible vertical transport of heat by the fluid flow---and concurrently very little radial velocity) is located at $\rov<r_0$. For the MHD case, $\rbcz=0.729\rsun$ and $\rov=0.710\rsun$. For the HD case, $\rbcz=0.726\rsun$ and $\rov=0.701\rsun$. 

\section{Non-dimensional parameters}\label{ap:nond}
The parameter regime of our simulations is described by four non-dimensional numbers (five for the MHD case) \citep{Hindman2020}: the flux Rayleigh number $\rm{Ra_F}$, the Ekman number $\rm{Ek}$, the dissipation number $\rm{Di}$, the thermal Prandtl number $\prt$, and (for the MHD case only) the magnetic Prandtl number $\prm$. The first four numbers are the same in both simulations. These numbers are defined and evaluated as
\begin{align*}
	{\rm{Ra_F}} &\equiv \frac{\tilde{g}\tilde{F}H^4}{c_{\rm{p}}\tilde{\rho}\tilde{T}\tilde{\nu}\tilde{\kappa}^2}=\sn{7.50}{5},\\
	{\rm{Ek}} &\equiv \frac{\tilde{\nu}}{\Omega_0H}=\sn{1.07}{-3},\\
	{\rm{Di}} &\equiv \frac{\tilde{g}H}{c_{\rm{p}}\tilde{T}}=1.72,\\
	{\rm{Pr}} &\equiv \frac{\tilde{\nu}}{\tilde{\kappa}}=1,\\
	\text{and}\ \ \ \ \ \prm &\equiv \frac{\tilde{\nu}}{\tilde{\eta}}=4.
\end{align*}
Here, we define the system's length-scale as $H\equiv\ro-r_0$, the tildes denote volume averages of the underlying reference-state radial profiles (from $r_0$ to $\ro$), and $F(r)\equiv(1/r^2)\int_{\ri}^r Q(x)x^2dx$ is approximately the energy that convection and conduction must carry to maintain a statistically steady state.

In each simulation's equilibrated state, several diagnostic non-dimensional numbers describe the system: the Reynolds number $\rm{Re}$, the magnetic Reynolds number $\rm{Re_m}$, the Rossby number $\rm{Ro}$, and the buoyancy parameter $\rm{B}$. These are defined as
\begin{align*}
	{\rm{Re}} &\equiv \frac{\tilde{v}^\prime H}{\tilde{\nu}},\\
	{\rm{Re_m}} &\equiv \frac{\tilde{v}^\prime H}{\tilde{\eta}},\\
	{\rm{Ro}} &\equiv \frac{\tilde{v}^\prime}{2\Omega_0H},\\
	\text{and}\ \ \ \ \ {\rm{B}} & \equiv \frac{\widetilde{N^2}}{\Omega_0^2}
\end{align*}
and evaluated in Table \ref{table:nond}. Here, the tildes denote combined temporal and volume averages, $v^\prime\equiv|\vecv^\prime|$, $N^2\equiv(g/c_{\rm{p}})[d\overline{S}/dr + \av{dS^\prime/dr}_{\rm{sph}}]$ is the squared buoyancy frequency, and $S^\prime$ is the entropy fluctuation from the background state.

\begin{table}
	\caption{{\bf{Diagnostic non-dimensional numbers for the HD and MHD cases}}. Diagnostic parameters defined in the text are shown volume-averaged over the convection zone, overshoot layer, and radiative interior.}
	\centering
	\begin{tabular}{l   l l l  |  l l l}
		\hline
		& & \bf{HD case} & & & \bf{MHD case}  & \\
		\hline
		& Convection zone &  Overshoot layer & Radiative interior & Convection zone &  Overshoot layer & Radiative interior\\ 
		\hline
		$\rm{Re}$ & $45.8$ & $58.4$ & $15.7$ & $36.3$ & $21.6$ & $2.86$\\
		\hline
		$\rm{Re_m}$ & - & - & - & $145$& $86.3$ & $11.5$ \\
		\hline
		$\rm{Ro}$ & $\sn{2.47}{-2}$ & $\sn{1.48}{-2}$ &$\sn{2.59}{-3}$ & $\sn{1.96}{-2}$ & $\sn{5.58}{-3}$ & $\sn{4.84}{-4}$\\
		\hline
		$\rm{B}$ & $-0.725$ & $1$,$130$ & $28$,$000$& $-0.710$ & $206$ & $26$,$700$\\
		\hline
	\end{tabular}
\label{table:nond}
\end{table}

	
	
\end{document}